\documentclass[10pt,letterpaper]{article}
\usepackage[margin=1in]{geometry}

\usepackage{ltexpprt}
\usepackage[hidelinks]{hyperref}

\usepackage{amsmath,amsfonts} 
\usepackage{algorithmic}
\usepackage{graphicx}
\usepackage{textcomp}
\usepackage{xcolor}

\usepackage{multirow} 
\usepackage{booktabs} 

\usepackage[normalem]{ulem}
\useunder{\uline}{\ul}{}

\usepackage{pgf}
\usepackage{caption}
\usepackage{subcaption}
\usepackage{enumitem}
\usepackage{stfloats}

\newtheorem{obs}{Observation}
\newtheorem{defn}{Definition}
\newtheorem{problem}{Problem}

\newcommand{\smallsection}[1]{{\vspace{0.05in} \noindent {\bf{\underline{\smash{#1}}}}}}

\def\BibTeX{{\rm B\kern-.05em{\sc i\kern-.025em b}\kern-.08em
    T\kern-.1667em\lower.7ex\hbox{E}\kern-.125emX}}
    
\begin{document}

\newcommand\relatedversion{}
\renewcommand\relatedversion{\thanks{The full version of the paper can be accessed at \protect\url{https://arxiv.org/abs/1902.09310}}} 

\title{On the Persistence of Higher-Order Interactions in Real-World Hypergraphs}
\author{Hyunjin Choo\thanks{School of Electrical Engineering, KAIST, Seoul, South Korea, choo@kaist.ac.kr}
\and Kijung Shin\thanks{Kim Jaechul Graduate School of AI and School of Electrical Engineering, KAIST, Seoul, South Korea, kijungs@kaist.ac.kr}}

\title{\Large On the Persistence of Higher-Order Interactions in Real-World Hypergraphs}
\author{Hyunjin Choo\thanks{School of Electrical Engineering, KAIST, Seoul, South Korea. choo@kaist.ac.kr}
\and Kijung Shin\thanks{Kim Jaechul Graduate School of AI and School of Electrical Engineering, KAIST, Seoul, South Korea. kijungs@kaist.ac.kr}}

\date{}

\maketitle







\begin{abstract}
A \textit{hypergraph}, which generalizes an ordinary graph, naturally represents group interactions as \textit{hyperedges} (i.e., arbitrary-sized subsets of nodes).
Such group interactions are ubiquitous: the sender and receivers of an email, the co-authors of a publication, and the items co-purchased by a customer, to name a few.
A \textit{higher-order interaction} (HOI) in a hypergraph is defined as the co-appearance of a set of nodes in any hyperedge. 
Our focus is the persistence of HOIs repeated over time, which is naturally interpreted as the strength of group relationships, aiming at answering three questions:
(a) How do HOIs in real-world hypergraphs persist over time?
(b) What are the key factors governing the persistence?
(c) How accurately can we predict the persistence?

In order to answer the questions above, we investigate the persistence of HOIs in 13 real-world hypergraphs from six domains.
First, we define how to measure the persistence of HOIs. Then, we examine global patterns and anomalies in the persistence, revealing a power-law relationship. 
After that, we study the relations between the persistence and 16 structural features of HOIs, some of which are closely related to the persistence.
Lastly, based on the 16 structural features, we assess the predictability of the persistence under various settings and find strong predictors.
Note that predicting the persistence of HOIs has many potential applications, such as recommending items to be purchased together and predicting missing recipients of emails.


\end{abstract}

    \section{Introduction}
    \label{sec:intro}
    A graph is a simple but powerful model for describing pairwise relationships or interactions between entities. It has been widely used for representing social networks (i.e., friendships between people), hyperlink networks (i.e., connections between web pages), purchase history (i.e., connections between a user and an item that she purchased), to name a few.
Real-world graphs have been examined extensively through descriptive and predictive analytics.
The former aims to discover structural and temporal patterns; and the latter focuses on predicting unknown or future states of graphs.
    


A \textit{hypergraph} is a generalization of a graph.
While an edge in ordinary graphs joins exactly two nodes, a \textit{hyperedge} in hypergraphs joins an arbitrary number of nodes.
This flexibility in the size of hyperedges allows a hypergraph to naturally represent group interactions among an arbitrary number of entities, such as the sender and the receivers of an email, the co-authors of a publication, and the items co-purchased by a customer.

A \textit{higher-order interaction} (HOI) in a hypergraph is defined as the co-appearance of a set of nodes in any hyperedge, and
it can be represented simply as a non-singleton subset of nodes.
For example, if four co-authors, $A$, $B$, $C$, and $D$, publish a paper together, any combination of them (i.e., any of $\{A, B\}$, $\{A, C\}$, $\{A, D\}$, $\{B, C\}$, $\{B, D\}$, $\{C, D\}$, $\{A, B, C\}$, $\{A, B, D\}$, $\{A, C, D\}$, $\{B, C,$ $D\}$, and $\{A, B, C, D\}$) becomes a HOI.
Note that HOIs can appear repeatedly over time. 
    
In this paper, we focus on the \textit{persistence} of HOIs repeated over time, which can naturally be used to measure the strength or robustness of group relations.
Moreover, predicting the persistence of HOIs has many potential applications, such as recommending groups (e.g., Facebook groups) in online social networks, recommending multiple items together, and predicting missing recipients of emails.
The significance of the persistence of HOIs brings up important questions that have not been answered yet:
(a) How do higher-order interactions in real-world hypergraphs persist over time?
(b) What are the key factors governing the persistence?
(c) How accurately can we predict the persistence?



In order to answer these questions, we empirically investigate the persistence of HOIs in 13 real-world hypergraphs from six domains.
We define the measure of the persistence of HOIs, and by using the measure, we closely examine the persistence at three different levels (hypergraphs, groups, and nodes), with a focus on patterns, predictability, and predictors.
\begin{itemize}[leftmargin=*]
    \setlength\itemsep{0em}
    \item \textbf{Patterns}: 
    We reveal power-laws in the persistence and examine how they vary depending on the size of HOIs.
    We explore relations between the persistence and $16$ group- or node-level structural features, and we find some interesting correlations (e.g., with entropy in the sizes of hyperedges including them).
    \item \textbf{Predictibility}: Based on the $16$ features, we assess the predictability of the future persistence of HOIs. We show how the predictability varies depending on the sizes of HOIs and how long we observe HOIs for.
    \item \textbf{Predictors}: We find strong group- and node-level predictors of the persistence of HOIs, through Gini importance-based feature selection. The strongest predictors are (a) the number of hyperedges containing the HOI and (b) the average (weighted) degree of the neighbors of each node in the HOIs.
\end{itemize}

\begin{figure}[t]
    \centering
    \vspace{-6mm}
    \includegraphics[width=0.6\linewidth]{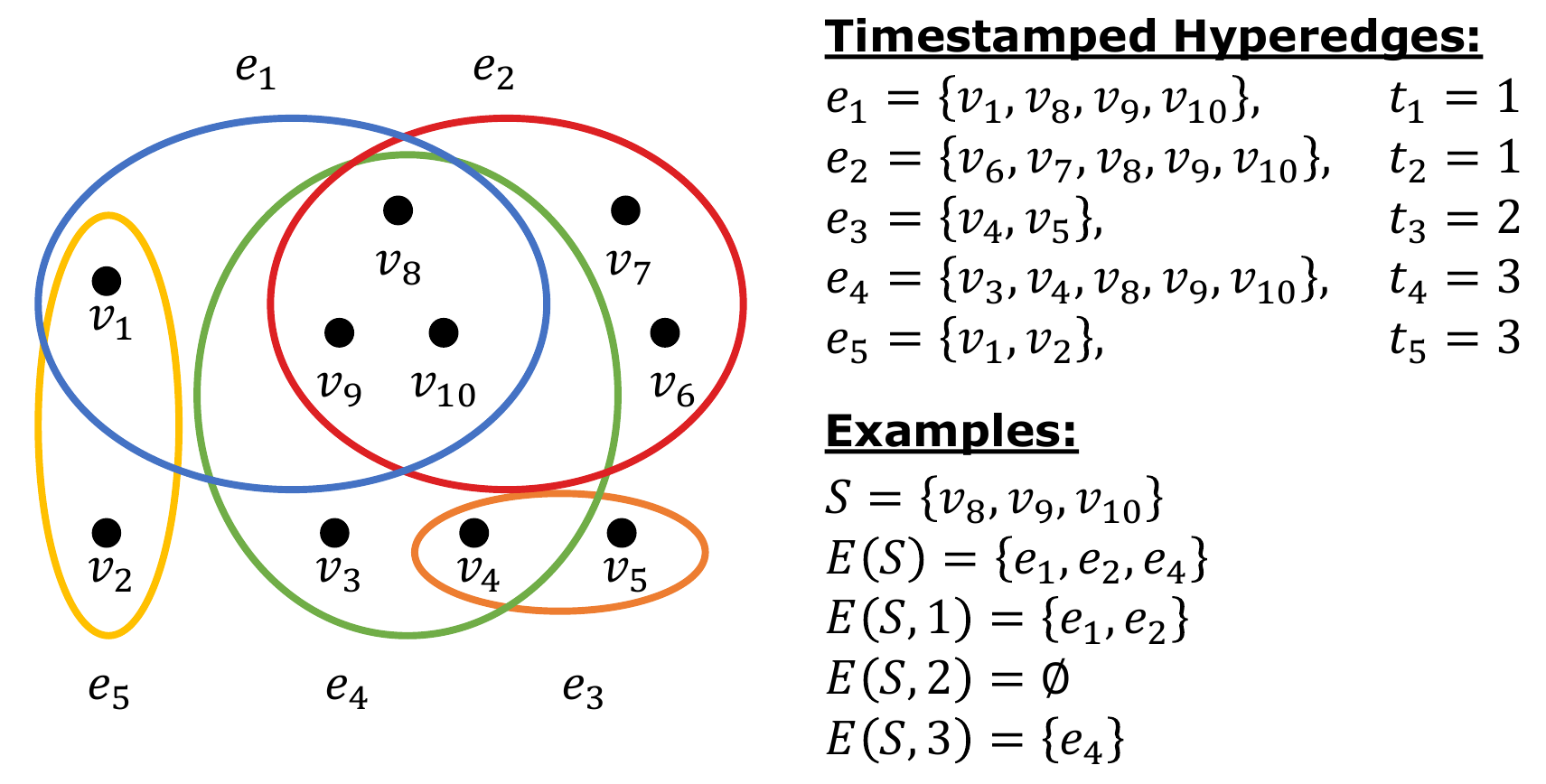}
    \caption{\label{fig:concept}\textbf{Examples of Concepts.}}
\end{figure}


In Section~\ref{sec:related}, we review some related studies. 
In Section~\ref{sec:prelim}, we give some preliminaries and describe the datasets used.
In Section~\ref{sec:obs}, we examine the persistence of HOIs.
In Section~\ref{sec:predict}, we assess	the predictability of the persistence of HOIs and the importance of features.
In Section~\ref{sec:summary}, we conclude our paper.

    \section{Related Work}
    \label{sec:related}
    In this section, we review empirical studies of structural and temporal properties of real-world (hyper)graphs.

\smallsection{Structural Properties.}
    There have been extensive studies on structural properties of static graphs or snapshots of dynamic graphs. 
    For example, their distance~\cite{albert1999diameter}, degree distribution~\cite{newman2001random}, and clustering coefficients~\cite{watts1998collective} were examined. 
    Regarding hypergraphs, 
    Do et al.~\cite{do2020structural} generalized clique expansion to represent relationships between subsets of nodes as graphs and analyzed their structures using the above graph metrics.
    Kook et al.~\cite{kook2020evolution} extended some graph metrics to hypergraphs and applied them directly to real-world hypergraphs. 
    Benson et al.~\cite{benson2018simplicial} focused on structural patterns related to simplicial closure events, 
    and Lee et al.~\cite{lee2021hyperedges} focused on those related to the overlaps of hyperedges.
    Lee et al.~\cite{lee2020hypergraph,lee2021thyme} extended network motifs to hypergraphs to investigate local connectivity patterns. 
	
\smallsection{Temporal Properties.}
   In addition to structural ones, temporal properties of dynamic graphs have received substantial attention \cite{leskovec2007graph,tang2010small,rossi2013modeling}.
   Leskovec et al. \cite{leskovec2007graph} showed that real-world graphs tend to densify with diameters shrinking over time, and the same patterns and shrinking intersections over time were observed in real-world hypergraphs \cite{kook2020evolution}.
   Benson et al.~\cite{benson2018simplicial} investigated the dynamics in the connectivity among triples of nodes. Specifically, they focused on pairwise relations between three nodes before a HOI among them (equivalently, a hyperedge that contains the three nodes) first appears, while our study focused on the repetition after its first appearance.
   Benson et al.~\cite{benson2018sequences} explored temporal patterns regarding the partial and full overlaps of hyperedges given as a sequence. Specifically, they showed the tendency of hyperedges to be more similar to recent hyperedges than more distant ones and reproduced the tendency through a model. 
   
    


\smallsection{Persistence of Pairwise Edges.}
    Hidalgo and Rodr{\'\i}guez-Sickert \cite{hidalgo2008dynamics} examined the relations between edge persistence and structural features (e.g., coreness and reciprocity) in a mobile phone network. 
	Belth et al.~\cite{belth2020mining} proposed a measure of the persistence of activity snippets (i.e., sequences of reoccurring edge-updates) and showed its usefulness in identifying anomalies. 
	
\smallsection{Relation to Our Work.} Our study systematically examines the persistence of HOIs in real-world hypergraphs, with a focus on patterns, predictability, and predictors, while past studies focused on other temporal aspects or the persistence of pairwise relations.


    \section{Preliminaries and Datasets}
    \label{sec:prelim}
    In this section, we first give some preliminaries and then describe the datasets used throughout this paper.
				
\begin{table}[t]
    \vspace{-5mm}
    \small
    \caption{\label{tab:notations}\textbf{Frequently-Used Notations.}}
    \centering
        \begin{tabular}{r|l}
        \toprule
        Notation            & Definition \\ \midrule
        $H = (V, E)$                 & Hypergraph with nodes $V$ and hyperedges $E$ \\
        $\bar{E}$                    & Set of high-order interactions (HOIs) in $H$ \\ \midrule
        $N(v)$                       & Set of nodes that co-appear with a node $v$ in any hyperedge \\
        $E(S)$                       & Set of hyperedges that contain a subset $S$ of $V$ \\
        $E(S,t)$                     & Set of hyperedges that contain a subset $S$ of $V$ at a time $t$ \\ 
        \midrule
        $H' = (V, E', \Omega)$        & Projected graph of $H$ \\
        $N'(v)$                       & Neighbors of a node $v$ in $H'$ \\
        \midrule
        $P(S,T)$                     & Persistence of a HOI $S$ over a time range $T$ \\ 
        $P_k(v,T)$                     & $k$-node persistence of a node $v$ over a time range $T$ \\ 
        \midrule
        MI & Mutual information \\
        CC & Correlation coefficient \\
        \bottomrule
        \end{tabular}
\end{table}

\subsection{Preliminary Concepts}
\label{sec:prelim:concepts}    
	    
Below, we introduce several preliminary concepts. See Table~\ref{tab:notations} for some frequently-used notations and Fig.~\ref{fig:concept} for examples.

\smallsection{Hypergraphs.}
Consider a \textit{hypergraph} $H=(V, E)$ where $V$ is the set of \textit{nodes} and $E \subseteq 2^V$ is the set of \textit{hyperedges}.
Each hyperedge $e_i$ is a subset of $V$, and we define the size $|e_i|$ of a hyperedge $e_i$ as the number of nodes in  $e_i$.
The neighbors $N(v):=\{u\in V: \exists e_i\in E \text{ s.t. } \{u,v\} \subseteq e_i \}\setminus\{v\}$ of each node $v$ is  the set of nodes included together with $v$ in any hyperedge in $E$.
		    
\smallsection{Higher-Order Interactions.}
A \textit{high-order interaction} (HOI) is defined as the co-appearance of a set of nodes in any hyperedge, and we represent it as a non-singleton subset of nodes.
That is, the set of HOIs in $H=(V,E)$ can be expressed as $\bar{E}:=\{S\subseteq V: |S|> 1 \text{ and } \exists e_i\in E \text{ s.t. } S \subseteq e_i  \}$.
For example, a hyperedge $\{A,B,C\}\in E$ implies four HOIs: $\{A, B\}$, $\{A, C\}$, $\{B, C\}$, or $\{A, B, C\}$.
For each HOI $S\in \bar{E}$, we use $E(S):=\{e_i\in E : S\subseteq e_i\}$ to denote the set of hyperedges containing $S$.
		    
\smallsection{Timestamped Hyperedges.}
Assume each hyperedge $e_i\in E$ is associated with the timestamp $t_i$. For each HOI $S\in \bar{E}$, we use $E(S,t)=\{e_i\in E(S) : t_i = t\}$ to denote the set of hyperedges at time $t$ containing $S$.
		
\smallsection{Projected Graphs.}
The \textit{projected graph} (a.k.a., \textit{clique expansion}) $H' = (V, E', \Omega)$ of a hypergraph $H=(V,E)$ is a pairwise graph where (a) any two nodes $u$ and $v$ are joined by an edge $\{u,v\}$ if and only if they co-appear in any hyperedge in $E$ and (b) the weight $\Omega(\{u,v\})$ of the edge equals the number of hyperedges where $u$ and $v$ co-appear. 
That is, $E':=\{\{u,v\}\in {V \choose 2}: \exists e\in E \text{ s.t. } \{u,v\}\subseteq e \}$, and $\Omega(\{u,v\}):=|\{e\in E:  \{u,v\}\subseteq e \}|$.
The projected graph of a hypergraph is obtained by replacing each hyperedge with the clique with the nodes in the hyperedge.
For each node $v$, we use $N'(v)$ to indicate the set of neighbors of $v$ in $H'$.

\subsection{Datasets}
\label{sec:prelim:data}
	
Throughout this study, we use the 13 real-world hypergraphs in Table~\ref{tab:datasets}.\footnote{\url{https://www.cs.cornell.edu/~arb/data/}} 
\begin{itemize}[leftmargin=*]
    \setlength\itemsep{0em}
    \item \textbf{Co-authorship (DBLP, Geology, \& History)}: Each node is an author of a publication. Each hyperedge is the set of authors of a publication.
    \item \textbf{Contact (High \& Primary)}: Each node is a person. Each hyperedge is a group interaction recorded by wearable sensors.
	\item \textbf{Email (Eu \& Enron)}: Each node is an email address. Each hyperedge is the set of the sender and all receivers of an email.
	\item \textbf{NDC-Classes}: Each node is a class label, and each hyperedge is the set of class labels applied to a drug.
	\item \textbf{NDC-Substances}: Each node is a substance. Each hyperedge is the set of substances in a drug.
    \item \textbf{Tags (Math.sx \& Ubuntu)}: Each node is a tag. Each hyperedge is the set of tags added to a question.
    \item \textbf{Threads (Math.sx \& Ubuntu)}: Each node is a user. Each hyperedge is the set of users who participate in a thread lasting for at most 24 hours.
\end{itemize}

\begin{table}[t]
    \vspace{-5mm}
    \small
	\caption{\label{tab:datasets}\textbf{Summary of datasets.}}
	\centering
        \begin{tabular}{llcccc}
        \toprule
        Domain               & Dataset & \# Nodes & \# Hyperedges & Time Range & Time Unit \\ \midrule
        \multirow{3}{*}{Coatuhorship} & DBLP             & 1,924,991         & 3,700,067              & 83                  & 1 Year               \\
                                      & Geology          & 1,256,385         & 1,590,335              & 219                 & 1 Year               \\
                                      & History          & 1,014,734         & 1,812,511              & 219                 & 1 Year               \\ \midrule
        \multirow{2}{*}{Contact}      & High             & 327               & 172,035                & 84                  & 1 Day                \\
                                      & Primary          & 242               & 106,879                & 108                 & 6 Hours               \\ \midrule
        \multirow{2}{*}{Email}        & Enron            & 143               & 10,883                 & 43                  & 1 Month              \\
                                      & Eu               & 998               & 234,760                & 38                  & 2 Weeks               \\ \midrule
        \multirow{2}{*}{NDC}          & Classes          & 1,161             & 49,724                 & 59                 & 2 Years               \\
                                      & Substances       & 5,311             & 112,405                & 59                 & 2 Years               \\ \midrule
        \multirow{2}{*}{Tags}         & Math.sx          & 1,629             & 822,059                & 89                  & 1 Month              \\
                                      & Ubuntu           & 3,029             & 271,233                & 104                 & 1 Month              \\ \midrule
        \multirow{2}{*}{Threads}      & Math.sx           & 176,445           & 719,792                & 85                  & 1 Month              \\
                                      & Ubuntu           & 125,602           & 192,947                & 92                  & 1 Month              \\ \bottomrule
		\end{tabular}
\end{table}

The time unit used for each dataset is stated in Table~\ref{tab:datasets}.
In each dataset, we use only the hyperedges that contain at most $25$ nodes to prevent extremely large hyperedges (e.g., a paper co-authored by more than 100 authors), which are very few in numbers, from significantly affecting the result of our analysis.
In \cite{appendix}, we also use randomized hypergraphs to compare results in them with those in the real-world hypergraphs.

    \section{Observations}
    \label{sec:obs}
    In this section, we examine the persistence of HOIs in the real-world hypergraphs.
First, we define how to measure the persistence of a HOI.
Then, we examine global and local patterns in the persistence of HOIs.

\subsection{Measures: Persistence \& Structural Features}
\label{sec:obs:defn}

    Below, we define the persistence of a high-order interaction (HOI). Then, we describe how we measure the persistence and structural features. 

    \smallsection{Definition.}
	We define the persistence of a HOI, aiming at measuring how steadily a HOI appears over a long period of time.
	That is, we aim to design the persistence measure so that short-lived HOIs have low persistence even if they are bursty, while long-lasting HOIs have high persistence even if they are infrequent.
	Thus, we consider coarse-grained discrete time units stated in Table~\ref{tab:datasets} and check whether each HOI appears at least once at each time unit, as formalized in Definition~\ref{defn:persistence}.
	
	\vspace{0.5mm}
	\noindent\fbox{%
    \parbox{0.98\columnwidth}{%
        \vspace{-3.5mm}
        \begin{defn}[Persistence]
	        \label{defn:persistence}
	    We define the \textit{persistence} $P(S,T)$ of a HOI $S\in \bar{E}$ over a time range $T$ as
	    \vspace{-3mm}
	    \begin{equation}
	        P(S,T) := \sum\nolimits_{t\in T} I(S,t) \text{,} \label{eq:persistence}
	        \vspace{-1mm}
        \end{equation}
        where $I(S,t)$ is $1$ if $S$ is included in any hyperedge at time $t$ (i.e., if $|E(S, t)| \geq 1$), and $0$ otherwise.
    \end{defn}
    \vspace{-3.5mm}
    }%
    }
	\vspace{0.2mm}
    
	For example in Fig.~\ref{fig:concept}, the persistence of $S$ over $T=[1,3]$ is $2$, i.e., $P(S,[1,3])=\sum_{t=1}^{3}I(S,t) = 1+0+1 = 2$.
	
	\begin{table}[t]
        \vspace{-5mm}
        \small
    	\centering
    	\caption{\label{tab:obs1_tab}
    	\textbf{Observations 1 and 2.}
    	The goodness-of-fit $R^2$ of  straight lines fitted on a log-log scale is high and often close to $1$.
    	The exponents of fitted power-law distributions and the average persistence decrease as the size of the HOI increases.}
    	\setlength{\tabcolsep}{12pt}
        \begin{tabular}{l|ccc|ccc|ccc}
        \toprule
        Dataset        & \multicolumn{3}{c|}{$R^2$ of  Fitted Line} & \multicolumn{3}{c|}{\begin{tabular}[c]{@{}c@{}}Power-Law Exponent\\ (Relative)\end{tabular}} & \multicolumn{3}{c}{\begin{tabular}[c]{@{}c@{}}Average Persistence\\ (Relative)\end{tabular}} \\ \midrule
        Size of HOIs      & 2                   & 3                   & 4                   & 2                                 & 3                                & 4                                & 2                                 & 3                                & 4                                \\ \midrule
        DBLP             & 0.98 & 0.99 & 0.96 & 1.00                              & 0.72                             & 0.57                             & 1.00                              & 0.76                             & 0.73                             \\
        Geology          & 0.97 & 0.99 & 0.95 & 1.00                              & 0.83                             & 0.67                             & 1.00                              & 0.83                             & 0.80                             \\
        History          & 0.96 & 0.92 & 0.99 & 1.00                              & 0.85                             & 0.55                             & 1.00                              & 0.95                             & 0.94                             \\
        High             & 0.95 & 0.99 & 1.00 & 1.00                              & 0.65                             & 0.51                             & 1.00                              & 0.55                             & 0.48                             \\
        Primary          & 0.85       & 0.93 & 0.99 & 1.00                              & 0.62                             & 0.64                             & 1.00                              & 0.40                             & 0.35                             \\
        Enron            & 0.80                & 0.90 & 0.95 & 1.00                              & 0.53                             & 0.42                             & 1.00                              & 0.51                             & 0.35                             \\
        Eu               & 0.90       & 0.88       & 0.86       & 1.00                              & 0.74                             & 0.67                             & 1.00                              & 0.75                             & 0.66                             \\
        Classes          & 0.50                & 0.36                & 0.27                & 1.00                              & 0.99                             & 0.99                             & 1.00                              & 1.00                             & 1.00                             \\
        Substances       & 0.93 & 0.90 & 0.87       & 1.00                              & 0.73                             & 0.59                             & 1.00                              & 0.68                             & 0.60                             \\
        Ubuntu (Tag)     & 0.99 & 0.99 & 0.97 & 1.00                              & 0.74                             & 0.66                             & 1.00                              & 0.55                             & 0.43                             \\
        Math.sx (Tag)    & 0.97 & 0.97 & 0.97 & 1.00                              & 0.65                             & 0.54                             & 1.00                              & 0.51                             & 0.34                             \\
        Ubuntu (Thr)      & 0.97 & 0.94 & -                   & 1.00                              & 0.51                             & -                                & 1.00                              & 0.96                             & -                                \\
        Math.sx (Thr)     & 0.98 & 0.98 & 1.00 & 1.00                              & 0.64                             & 0.24                             & 1.00                              & 0.83                             & 0.83                             \\ \midrule
        \textbf{Average} & \textbf{0.90} & \textbf{0.90} & \textbf{0.90} & \textbf{1.00}                     & \textbf{0.71}                    & \textbf{0.59}                    & \textbf{1.00}                     & \textbf{0.72}                    & \textbf{0.63}                    \\ \bottomrule
        \multicolumn{10}{l}{-: not enough HOIs.}
        \end{tabular}
	\end{table}
	
%
	
	\smallsection{How to Measure.}
	We divide the time units for measuring structural features and the persistence.
	For each HOI, we compute its structural features from all hyperedges appearing over the next $T_{s}$ time units after its first appearance.
    Then, we measure its persistence by Eq.~\eqref{eq:persistence} over the next $T_{p}$ time units.
    That is, our process for each HOI $S$ is as follows:
	(1) $S$ appears in a hyperedge for the first time at time $t$. 
	(2) We compute its structural features using only the hyperedges appearing between time $t+1$ and $t+T_s$.
	(3) We measure its persistence between time $t+T_s+1$ and $t+T_s+T_p$.
	According to our preliminary study, our findings are insensitive to the values of $T_{s}$ and $T_{p}$.
	Thus, we assume $T_{s}=5$ and $T_{p}=10$, unless otherwise stated.
	
	In order to examine the relation between persistence and each structural feature, we measure the \textit{Pearson correlation coefficient} (CC) and \textit{normalized mutual information} (MI)\footnote{Normalized mutual information is a normalized measure of the mutual information score that scales from 0 (no mutual information) to 1 (perfect correlation).} between them.

	

        
		

	\begin{figure*}[t]
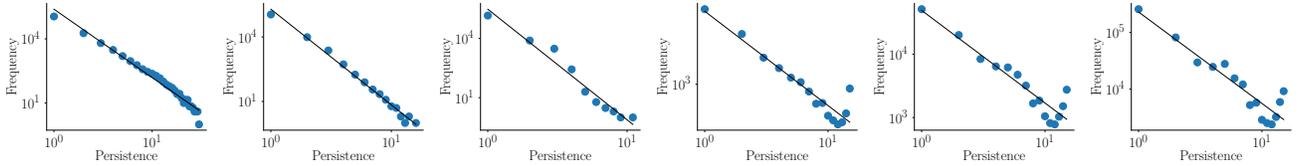

	    \vspace{-2mm}
	    \centering
		\begin{subfigure}{.16\textwidth}
			\resizebox{\textwidth}{!}{\input{./figure/coauth-DBLP_i_83_30_25_scatter_2.pgf}}
			\caption{\footnotesize DBLP ($|S|=2$)} \label{fig:obs1_fig_1}
		\end{subfigure}
		\begin{subfigure}{.16\textwidth}
			\resizebox{\textwidth}{!}{\input{./figure/coauth-DBLP_i_83_30_25_scatter_3.pgf}}
			\caption{\footnotesize DBLP ($|S|=3$)} \label{fig:obs1_fig_2}
		\end{subfigure}
		\begin{subfigure}{.16\textwidth}
			\resizebox{\textwidth}{!}{\input{./figure/coauth-DBLP_i_83_30_25_scatter_4.pgf}}
			\caption{\footnotesize DBLP ($|S|=4$)} \label{fig:obs1_fig_3}
		\end{subfigure}
		\begin{subfigure}{.16\textwidth}
			\resizebox{\textwidth}{!}{\input{./figure/email-Eu_i_38_15_25_scatter_2.pgf}}
			\caption{\footnotesize Eu ($|S|=2$)} \label{fig:obs1_fig_4}
		\end{subfigure}
		\begin{subfigure}{.16\textwidth}
			\resizebox{\textwidth}{!}{\input{./figure/email-Eu_i_38_15_25_scatter_3.pgf}}
			\caption{\footnotesize Eu ($|S|=3$)} \label{fig:obs1_fig_5}
		\end{subfigure}
		\begin{subfigure}{.16\textwidth}
			\resizebox{\textwidth}{!}{\input{./figure/email-Eu_i_38_15_25_scatter_4.pgf}}
			\caption{\footnotesize Eu ($|S|=4$)} \label{fig:obs1_fig_6}
		\end{subfigure}
		\caption{\label{fig:obs1_fig}\textbf{Power-Laws in the Persistence of HOIs} Note the surprising abundance of highly persistent HOIs in the Eu (Email) dataset. See \cite{appendix} for results in the other datasets.}
	\end{figure*}

\subsection{Global Analysis: Persistence vs. Frequency}
\label{sec:obs:hypergraph}
    

    Below, we examine global patterns in the persistence of HOIs in the considered real-world hypergraphs.
	First, we measure the persistence of each HOI by Eq.~\eqref{eq:persistence} over the $30$ time units after its first appearance.\footnote{We use 15 and 20 time units for the NDC and Email datasets since they have a small number of time units.}
	Then, we compute the distributions of the persistence of HOIs of each size: $2$ (pairs), $3$ (triples), and $4$ (quadruples).
	After that, we fit a straight line to each distribution on a log-log scale.
	The goodness-of-fit $R^2$ is reported in Table~\ref{tab:obs1_tab}, and in all datasets, except the Enron, Eu, and Substances datasets, $R^2$ is over $0.9$ and often close to $1$. 

    \vspace{0.5mm}
	\noindent\fbox{%
    \parbox{0.98\columnwidth}{%
         \vspace{-3.5mm}
            \begin{obs}[Power-Laws in Persistence] \label{obs:powerlaw}
                In real-world hypergraphs, the persistence of HOIs tends to follow a power-law.
            \end{obs}
         \vspace{-3.5mm}
    }%
    }
    \vspace{0.2mm}
		
	The distributions of the persistence of HOIs in the DBLP (Co-authorship) and Eu (Email) datasets are shown in Fig.~\ref{fig:obs1_fig}.
	While the distributions from the DBLP dataset clearly obey power-laws, there exist anomalies that deviate from the fitted lines in the distributions from the Eu dataset.
	The anomalies indicate the surprising abundance of highly persistent HOIs. 
		
	Next, we compute how (a) the exponents (i.e.,  $k$ in $f(x)=ax^{-k}$) of the fitted power-law distributions and (b) the average persistence of HOIs of size 2, 3, or 4 change as the sizes of HOIs grow.
	In Table~\ref{tab:obs1_tab},
	we report the values relative to those when the size of the HOI is two. 
	In all datasets, except for the Primary dataset, the exponents and the average persistence decline as the size of the HOI grows. 
	
\vspace{0.5mm}
	\noindent\fbox{%
    \parbox{0.98\columnwidth}{%
         \vspace{-3.5mm}
        \begin{obs}[Size of HOIs] \label{obs:size}
	        In real-world hypergraphs, as HOIs grow in size, their average persistence and the exponents of fitted power-law distributions tend to decrease. 
	    \end{obs}
	     \vspace{-3.5mm}
    }%
    }

    




\subsection{Local Analysis (1): Group Features vs. Group Persistence}
\label{sec:obs:group}

    Below, we define eight structural features of a HOI as a group.
    Then, we examine the relations between these group features and the persistence of HOIs (i.e., group persistence).

    \smallsection{Group Features.}
    We consider the basic features of each HOI $S$ in Table~\ref{tab:features_HOI}. After a preliminary study, by combining these basic features, we measure the following eight structural features: (a) $\#$, (b) $\#/\cup$, (c) $\Sigma/(\Sigma\cup)$, (d) $\cap$, (e) $\#/\cap$, (f) $\Sigma/\cap$, (g) $\Sigma/\#$, and (h) $\mathcal{H}$.
    Note that $\#/\cup$ is the \textit{density} \cite{hu2017maintaining} of $E(S)$ 
    and $\Sigma/\#$ is the average size of the hyperedges containing $S$.
    
    \begin{table}[t]
    \small
    \vspace{-5mm}
    \caption{\label{tab:features_HOI}\textbf{Basic Features of Each HOI $S$.}}
    \centering
        \begin{tabular}{cll}
        \toprule
Symbol                  & Definition                                                     & Description                                                                                     \\ \midrule
\textbf{${\#}$}         & $|E(S)|$                                                       & number of hyperedges including $S$                   \\
\textbf{${\Sigma}$}     & $\sum_{e \in E(S)} |e|$                               & sum of the sizes of the hyperedges containing $S$    \\
\textbf{${\cup}$}       & $|\bigcup_{v \in S} E(\{v\})|$                        & number of hyperedges overlapping $S$                  \\
\textbf{${\Sigma\cup}$} & $\sum_{e \in \bigcup\nolimits_{v \in S} E(\{v\})}|e|$ & sum of the sizes of the hyperedges overlapping $S$    \\
\textbf{${\cap}$}       & $|\bigcap\nolimits_{v\in S} N(v)|$                             & number of common neighbors of $S$                     \\
\textbf{$\mathcal{H}$}  & {\small $Entropy([|e|:e\in E(S)])^*$}                                  & entropy in the sizes of the hyperedges containing $S$ \\ \bottomrule
\multicolumn{3}{l}{$*$ $[\cdot]$ indicates a multiset} \\ 
\end{tabular}
\end{table}

\begin{figure*}[t]
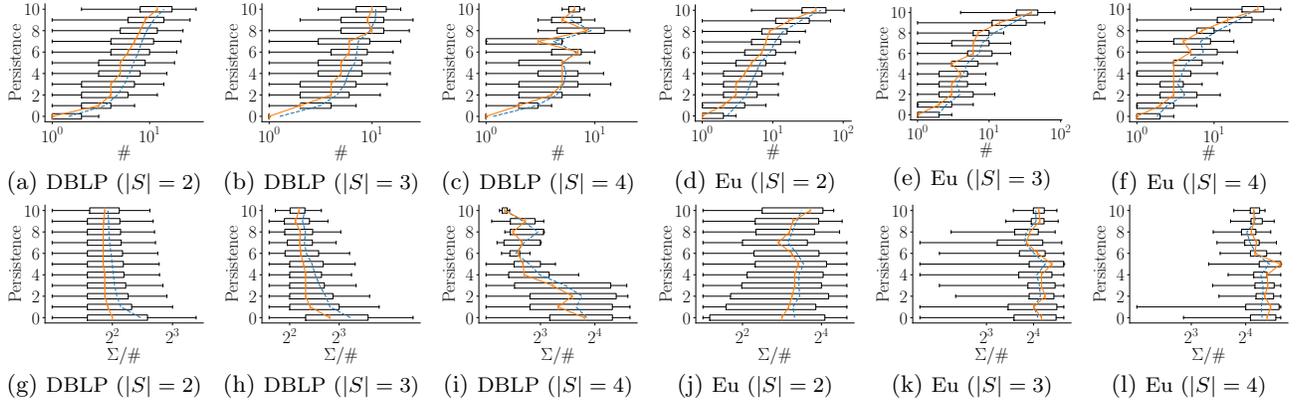

        \vspace{-6mm}
	    \centering
		\begin{subfigure}{.16\textwidth}
			\resizebox{\textwidth}{!}{\input{./figure/coauth-DBLP_h_p_b_83_10_5_25_Co-Occurrence-Count_2.pgf}}  
			\vspace{-6mm}
			\caption{\footnotesize DBLP ($|S|=2$)}
		\end{subfigure}
		\begin{subfigure}{.16\textwidth}
			\resizebox{\textwidth}{!}{\input{./figure/coauth-DBLP_h_p_b_83_10_5_25_Co-Occurrence-Count_3.pgf}}
			\vspace{-6mm}
			\caption{\footnotesize DBLP ($|S|=3$)}
		\end{subfigure}
		\begin{subfigure}{.16\textwidth}
			\resizebox{\textwidth}{!}{\input{./figure/coauth-DBLP_h_p_b_83_10_5_25_Co-Occurrence-Count_4.pgf}}
			\vspace{-6mm}
			\caption{\footnotesize DBLP ($|S|=4$)}
		\end{subfigure}
		\begin{subfigure}{.16\textwidth}
			\resizebox{\textwidth}{!}{\input{./figure/email-Eu_h_p_b_38_10_5_25_Co-Occurrence-Count_2.pgf}}
			\vspace{-6mm}
			\caption{\footnotesize Eu ($|S|=2$)}
		\end{subfigure}
		\begin{subfigure}{.16\textwidth}
			\resizebox{\textwidth}{!}{\input{./figure/email-Eu_h_p_b_38_10_5_25_Co-Occurrence-Count_3.pgf}}
			\vspace{-6mm}
			\caption{\footnotesize Eu ($|S|=3$)}
		\end{subfigure}
		\begin{subfigure}{.16\textwidth}
			\resizebox{\textwidth}{!}{\input{./figure/email-Eu_h_p_b_38_10_5_25_Co-Occurrence-Count_4.pgf}}
			\vspace{-6mm}
			\caption{\footnotesize Eu ($|S|=4$)}
		\end{subfigure}
	    \begin{subfigure}{.16\textwidth}
			\resizebox{\textwidth}{!}{\input{./figure/coauth-DBLP_h_p_b_83_10_5_25_HE-Size-Average_2.pgf}}
			\vspace{-6mm}
			\caption{\footnotesize DBLP ($|S|=2$)}
		\end{subfigure}
		\begin{subfigure}{.16\textwidth}
			\resizebox{\textwidth}{!}{\input{./figure/coauth-DBLP_h_p_b_83_10_5_25_HE-Size-Average_3.pgf}}
			\vspace{-6mm}
			\caption{\footnotesize DBLP ($|S|=3$)}
		\end{subfigure}
		\begin{subfigure}{.16\textwidth}
			\resizebox{\textwidth}{!}{\input{./figure/coauth-DBLP_h_p_b_83_10_5_25_HE-Size-Average_4.pgf}}
			\vspace{-6mm}
			\caption{\footnotesize DBLP ($|S|=4$)}
		\end{subfigure}
		\begin{subfigure}{.16\textwidth}
			\resizebox{\textwidth}{!}{\input{./figure/email-Eu_h_p_b_38_10_5_25_HE-Size-Average_2.pgf}}
			\vspace{-6mm}
			\caption{\footnotesize Eu ($|S|=2$)}
		\end{subfigure}
		\begin{subfigure}{.16\textwidth}
			\resizebox{\textwidth}{!}{\input{./figure/email-Eu_h_p_b_38_10_5_25_HE-Size-Average_3.pgf}}
			\vspace{-6mm}
			\caption{\footnotesize Eu ($|S|=3$)}
		\end{subfigure}
		\begin{subfigure}{.16\textwidth}
			\resizebox{\textwidth}{!}{\input{./figure/email-Eu_h_p_b_38_10_5_25_HE-Size-Average_4.pgf}}
			\vspace{-6mm}
			\caption{\footnotesize Eu ($|S|=4$)}
		\end{subfigure}
		\caption{\label{fig:group_fig}\textbf{Group Features vs. Group Persistence.} 
		The distribution of $\#$ and $\Sigma/\#$  (i.e., the number and average size of hyperedges containing each HOI) of HOIs with each level of persistence in two datasets. \textcolor{blue}{Blue} lines indicate means and \textcolor{orange}{orange} lines indicate medians. See \cite{appendix} for results in the other datasets.} 
	\end{figure*}

    \smallsection{Observations.}
	The mutual information (MI) and Pearson correlation coefficients (CC) between each structural group feature and the persistence are summarized in Table~\ref{tab:feature_persistence_tab}.
	Most features are positively correlated with the persistence, and on average, the CC is strongest for $\#$,
	(i.e., the number of hyperedges containing each HOI $S$), 
	followed by $\mathcal{H}$ (i.e., the entropy in the sizes of hyperedges containing each HOI $S$), 
	and then $\Sigma/\cap$.
	Notably, $\Sigma/\#$  (i.e., the average size of the hyperedges containing each HOI $S$) is the only feature that is negatively correlated with the persistence.
	We show in Fig.~\ref{fig:group_fig} the distributions of $\#$ and $\Sigma/\#$ of HOIs with each level of persistence in two datasets.
    
\vspace{0.5mm}
	\noindent\fbox{%
    \parbox{0.98\columnwidth}{%
        \vspace{-3.5mm}
        \begin{obs}\label{obs:group_group}
	        \textsc{(Group Features vs. Group Persistence)}
	        In real-world hypergraphs, the persistence of each HOI $S$ is positively correlated with (a) the number of hyperedges containing $S$ and (b) the entropy in the sizes of hyperedges containing $S$.
	    \end{obs}
	    \vspace{-3.5mm}
    }%
    }
	

    \begin{table*}[t]
        \vspace{-2mm}
    	\centering
    	\caption{\label{tab:feature_persistence_tab}\textbf{Features vs. Persistence.} Mutual information (MI) and correlation coefficients (CC) are averaged over all 13 datasets.
    	In each case, the first and second most strongly correlated features are in \textbf{bold} and \underline{\smash{underlined}}, respectively.
    	See \cite{appendix} for the results in each dataset.}
        \setlength{\tabcolsep}{2pt}
        \scalebox{0.76}{
        \begin{tabular}{c|c|cccccccc|cccccccc|cccccccc}
        \toprule
\multicolumn{1}{l|}{} & \multicolumn{1}{l|}{}                                  & \multicolumn{8}{c|}{Group Features vs. Group Persistence}                                                                                                                                                                 & \multicolumn{8}{c|}{Node Features vs. Group Persistence}                              & \multicolumn{8}{c}{Node Features vs. Node Persistence}                                   \\ \midrule
\textbf{}             & \begin{tabular}[c]{@{}c@{}}Size\\ of HOIs\end{tabular} & \textbf{$\#$} & \textbf{$\frac{\#}{\cup}$} & \textbf{$\frac{\Sigma}{\Sigma\cup}$} & \textbf{$\cap$} & \textbf{$\frac{\#}{\cap}$} & \textbf{$\frac{\Sigma}{\cap}$} & \textbf{$\frac{\Sigma}{\#}$} & \textbf{$\mathcal{H}$} & $d$   & $w$  & $c$   & $r$           & $\bar{d}$      & $\bar{w}$      & $l$   & $o$  & $d$  & $w$        & $c$  & $r$           & $\bar{d}$ & $\bar{w}$  & $l$   & $o$           \\ \midrule
\multirow{4}{*}{MI}   & 2                                                      & 0.13          & 0.11                       & {\ul 0.14}                           & 0.05            & 0.10                       & 0.12                           & 0.10                         & \textbf{0.15}          & 0.04  & 0.09 & 0.04  & \textbf{0.17} & 0.16           & {\ul 0.17}     & 0.15  & 0.08 & 0.35 & 0.43       & 0.28 & \textbf{0.53} & 0.49      & {\ul 0.51} & 0.43  & 0.41          \\
                      & 3                                                      & {\ul 0.11}    & 0.06                       & 0.08                                 & 0.05            & 0.08                       & 0.09                           & 0.08                         & \textbf{0.12}          & 0.03  & 0.06 & 0.04  & {\ul 0.09}    & 0.09           & \textbf{0.10}  & 0.09  & 0.05 & 0.30 & 0.37       & 0.24 & \textbf{0.44} & 0.42      & {\ul 0.44} & 0.37  & 0.34          \\
                      & 4                                                      & {\ul 0.11}    & 0.05                       & 0.07                                 & 0.06            & 0.07                       & 0.10                           & 0.07                         & \textbf{0.12}          & 0.03  & 0.05 & 0.06  & {\ul 0.07}    & 0.07           & \textbf{0.07}  & 0.07  & 0.04 & 0.26 & 0.31       & 0.21 & \textbf{0.36} & 0.35      & {\ul 0.36} & 0.31  & 0.30          \\ \cmidrule(l){2-26} 
                      & Avg.                                                   & {\ul 0.12}    & 0.08                       & 0.10                                 & 0.05            & 0.08                       & 0.11                           & 0.08                         & \textbf{0.13}          & 0.04  & 0.07 & 0.05  & {\ul 0.11}    & 0.11           & \textbf{0.11}  & 0.10  & 0.05 & 0.30 & 0.37       & 0.24 & \textbf{0.44} & 0.42      & {\ul 0.43} & 0.37  & 0.35          \\ \midrule
\multirow{4}{*}{CC}   & 2                                                      & \textbf{0.36} & 0.09                       & 0.09                                 & 0.17            & 0.19                       & 0.26                           & -0.08                        & {\ul 0.32}             & 0.05  & 0.09 & -0.01 & 0.07          & {\ul -0.12}    & \textbf{-0.14} & -0.08 & 0.09 & 0.15 & {\ul 0.22} & 0.14 & 0.08          & 0.00      & -0.07      & -0.02 & \textbf{0.26} \\
                      & 3                                                      & \textbf{0.31} & 0.10                       & 0.10                                 & 0.05            & 0.16                       & 0.20                           & -0.09                        & {\ul 0.25}             & -0.02 & 0.06 & -0.05 & 0.03          & {\ul -0.11}    & \textbf{-0.12} & -0.02 & 0.05 & 0.04 & {\ul 0.16} & 0.04 & 0.03          & -0.04     & -0.08      & -0.04 & \textbf{0.17} \\
                      & 4                                                      & \textbf{0.30} & 0.13                       & 0.13                                 & -0.01           & 0.17                       & 0.20                           & -0.10                        & {\ul 0.24}             & -0.07 & 0.03 & -0.09 & 0.03          & \textbf{-0.14} & {\ul -0.14}    & 0.03  & 0.00 & 0.03 & {\ul 0.12} & 0.01 & 0.02          & -0.05     & -0.07      & -0.04 & \textbf{0.13} \\ \cmidrule(l){2-26} 
                      & Avg.                                                   & \textbf{0.32} & 0.10                       & 0.11                                 & 0.07            & 0.17                       & 0.22                           & -0.09                        & {\ul 0.27}             & -0.01 & 0.06 & -0.05 & 0.04          & {\ul -0.12}    & \textbf{-0.13} & -0.02 & 0.05 & 0.07 & {\ul 0.17} & 0.06 & 0.04          & -0.03     & -0.07      & -0.03 & \textbf{0.19} \\ \bottomrule
        \end{tabular}
        }
	\end{table*}

\subsection{Local Analysis (2): Node Features vs. Group Persistence}
\label{sec:obs:node_group}
    
    Below, we examine the relations between the persistence of each HOI (i.e., group persistence) and the eight structural features of individual nodes involved in the HOI. 
    
    \smallsection{Node Features.}
    As described in Section~\ref{sec:obs:defn}, for each HOI appearing for the first time at time $t$, we consider the hypergraph $H$ consisting of all hyperedges appearing between time $t+1$ and $t+T_s$.
    As the structural features of each node $v$, we consider its (a) \textbf{degree} $d(v)$, (b) \textbf{weighted degree} $w(v)$, (c) \textbf{core number}~\cite{batagelj2003m} $c(v)$, (d) \textbf{PageRank} (the damping factor is set to $0.85$) ~\cite{page1999pagerank} $r(v)$, 
    (e) \textbf{average degree of neighbors} $\bar{d}(v)$, (f) \textbf{average weighted degree of neighbors} $\bar{w}(v)$, (g) \textbf{local clustering coefficient}~\cite{watts1998collective} $l(v)$.
    Additionally, we consider the (h) \textbf{number of occurrences} $o(v)$ of $v$ in $H$.
	
\begin{figure*}[t]
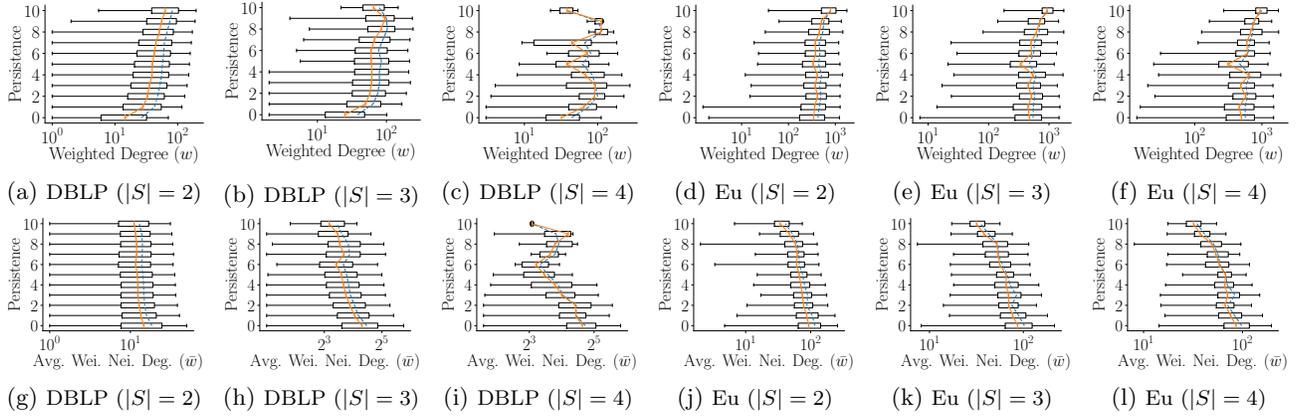

        \vspace{-3.5mm}
	    \centering
		\begin{subfigure}{.16\textwidth}
			\resizebox{\textwidth}{!}{\input{./figure/coauth-DBLP_p_p_b_83_10_5_25_Degree-Weighted_2.pgf}}  
			\vspace{-5mm}
			\caption{\footnotesize DBLP ($|S|=2$)}
		\end{subfigure}
		\begin{subfigure}{.16\textwidth}
			\resizebox{\textwidth}{!}{\input{./figure/coauth-DBLP_p_p_b_83_10_5_25_Degree-Weighted_3.pgf}}
			\caption{\footnotesize DBLP ($|S|=3$)}
		\end{subfigure}
		\begin{subfigure}{.16\textwidth}
			\resizebox{\textwidth}{!}{\input{./figure/coauth-DBLP_p_p_b_83_10_5_25_Degree-Weighted_4.pgf}}
			\vspace{-5mm}
			\caption{\footnotesize DBLP ($|S|=4$)}
		\end{subfigure}
		\begin{subfigure}{.16\textwidth}
			\resizebox{\textwidth}{!}{\input{./figure/email-Eu_p_p_b_38_10_5_25_Degree-Weighted_2.pgf}}
			\vspace{-5mm}
			\caption{\footnotesize Eu ($|S|=2$)}
		\end{subfigure}
		\begin{subfigure}{.16\textwidth}
			\resizebox{\textwidth}{!}{\input{./figure/email-Eu_p_p_b_38_10_5_25_Degree-Weighted_3.pgf}}
			\vspace{-5mm}
			\caption{\footnotesize Eu ($|S|=3$)}
		\end{subfigure}
		\begin{subfigure}{.16\textwidth}
			\resizebox{\textwidth}{!}{\input{./figure/email-Eu_p_p_b_38_10_5_25_Degree-Weighted_4.pgf}}
			\vspace{-5mm}
			\caption{\footnotesize Eu ($|S|=4$)}
		\end{subfigure}
		\begin{subfigure}{.16\textwidth}
			\resizebox{\textwidth}{!}{\input{./figure/coauth-DBLP_p_p_b_83_10_5_25_Avg-Nbr-Degree-Weighted_2.pgf}}
			\vspace{-5mm}
			\caption{\footnotesize DBLP ($|S|=2$)}
		\end{subfigure}
		\begin{subfigure}{.16\textwidth}
			\resizebox{\textwidth}{!}{\input{./figure/coauth-DBLP_p_p_b_83_10_5_25_Avg-Nbr-Degree-Weighted_3.pgf}}
			\vspace{-5mm}
			\caption{\footnotesize DBLP ($|S|=3$)}
		\end{subfigure}
		\begin{subfigure}{.16\textwidth}
			\resizebox{\textwidth}{!}{\input{./figure/coauth-DBLP_p_p_b_83_10_5_25_Avg-Nbr-Degree-Weighted_4.pgf}}
			\vspace{-5mm}
			\caption{\footnotesize DBLP ($|S|=4$)}
		\end{subfigure}
		\begin{subfigure}{.16\textwidth}
			\resizebox{\textwidth}{!}{\input{./figure/email-Eu_p_p_b_38_10_5_25_Avg-Nbr-Degree-Weighted_2.pgf}}
			\vspace{-5mm}
			\caption{\footnotesize Eu ($|S|=2$)}
		\end{subfigure}
		\begin{subfigure}{.16\textwidth}
			\resizebox{\textwidth}{!}{\input{./figure/email-Eu_p_p_b_38_10_5_25_Avg-Nbr-Degree-Weighted_3.pgf}}
			\vspace{-5mm}
			\caption{\footnotesize Eu ($|S|=3$)}
		\end{subfigure}
		\begin{subfigure}{.16\textwidth}
			\resizebox{\textwidth}{!}{\input{./figure/email-Eu_p_p_b_38_10_5_25_Avg-Nbr-Degree-Weighted_4.pgf}}
			\vspace{-5mm}
			\caption{\footnotesize Eu ($|S|=4$)}
		\end{subfigure}
		\caption{\label{fig:node_group_fig}\textbf{Node Features vs. Group Persistence.} 
		The distribution of averaged $w$ (i.e., weighted degree) and $\bar{w}$ (i.e., the average weighted degree of neighbors) of HOIs with each level of persistence in two datasets. \textcolor{blue}{Blue} lines indicate means and \textcolor{orange}{orange} lines indicate medians. See \cite{appendix} for results in the other datasets.}
	\end{figure*}

	\smallsection{Observations.}
	The mutual information (MI) and Pearson correlation coefficients (CC) between each structural node feature, which is averaged over the nodes involved in each HOI, and the persistence are summarized in Table~\ref{tab:feature_persistence_tab}.
	On average, the MI is largest for $\bar{w}$ (i.e., the average weighted degree of neighbors), $\bar{d}$ (i.e., the average degree of neighbors), and $r$ (i.e., PageRank).
	Notably, $\bar{w}$ and $\bar{d}$ are negatively correlated with the persistence.
	In addition to $r$, $w$ (i.e., weighted degree), and $o$ (i.e., the number of occurrences) are positively correlated with the persistence.
    The distributions of averaged $w$ and $\bar{w}$ of HOIs with each level of persistence in two datasets are shown in Fig.~\ref{fig:node_group_fig}.
    
    \vspace{0.5mm}
	\noindent\fbox{%
    \parbox{0.98\columnwidth}{%
    \vspace{-3.5mm}
        \begin{obs}\label{obs:node_group}
            \textsc{(Node Features vs. Group Persistence)}
	        In real-world hypergraphs, the persistence of each HOI is negatively correlated with the average (weighted) degree of neighbors of each node involved in the HOI.
	    \end{obs}
	    \vspace{-3.5mm}
    }%
    }
    \vspace{-1mm}
    

	\begin{figure*}[t]
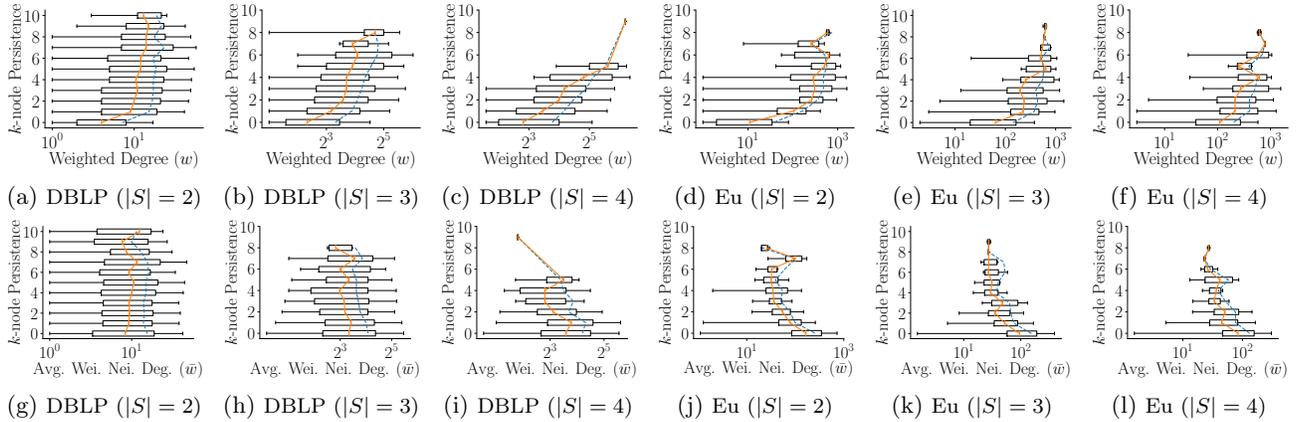

	    \vspace{-6mm}
	    \centering
		\begin{subfigure}{.16\textwidth}
			\resizebox{\textwidth}{!}{\input{./figure/coauth-DBLP_p_n_b_83_10_5_25_Degree-Weighted_2.pgf}}  
			\vspace{-5mm}
			\caption{\footnotesize DBLP ($|S|=2$)}
		\end{subfigure}
		\begin{subfigure}{.16\textwidth}
			\resizebox{\textwidth}{!}{\input{./figure/coauth-DBLP_p_n_b_83_10_5_25_Degree-Weighted_3.pgf}}
			\vspace{-5mm}
			\caption{\footnotesize DBLP ($|S|=3$)}
		\end{subfigure}
		\begin{subfigure}{.16\textwidth}
			\resizebox{\textwidth}{!}{\input{./figure/coauth-DBLP_p_n_b_83_10_5_25_Degree-Weighted_4.pgf}}
			\vspace{-5mm}
			\caption{\footnotesize DBLP ($|S|=4$)}
		\end{subfigure}
		\begin{subfigure}{.16\textwidth}
			\resizebox{\textwidth}{!}{\input{./figure/email-Eu_p_n_b_38_10_5_25_Degree-Weighted_2.pgf}}
			\vspace{-5mm}
			\caption{\footnotesize Eu ($|S|=2$)}
		\end{subfigure}
		\begin{subfigure}{.16\textwidth}
			\resizebox{\textwidth}{!}{\input{./figure/email-Eu_p_n_b_38_10_5_25_Degree-Weighted_3.pgf}}
			\vspace{-5mm}
			\caption{\footnotesize Eu ($|S|=3$)}
		\end{subfigure}
		\begin{subfigure}{.16\textwidth}
			\resizebox{\textwidth}{!}{\input{./figure/email-Eu_p_n_b_38_10_5_25_Degree-Weighted_4.pgf}}
			\vspace{-5mm}
			\caption{\footnotesize Eu ($|S|=4$)}
		\end{subfigure}
		\begin{subfigure}{.16\textwidth}
			\resizebox{\textwidth}{!}{\input{./figure/coauth-DBLP_p_n_b_83_10_5_25_Avg-Nbr-Degree-Weighted_2.pgf}}
			\vspace{-5mm}
			\caption{\footnotesize DBLP ($|S|=2$)}
		\end{subfigure}
		\begin{subfigure}{.16\textwidth}
			\resizebox{\textwidth}{!}{\input{./figure/coauth-DBLP_p_n_b_83_10_5_25_Avg-Nbr-Degree-Weighted_3.pgf}}
			\vspace{-5mm}
			\caption{\footnotesize DBLP ($|S|=3$)}
		\end{subfigure}
		\begin{subfigure}{.16\textwidth}
			\resizebox{\textwidth}{!}{\input{./figure/coauth-DBLP_p_n_b_83_10_5_25_Avg-Nbr-Degree-Weighted_4.pgf}}
			\vspace{-5mm}
			\caption{\footnotesize DBLP ($|S|=4$)}
		\end{subfigure}
		\begin{subfigure}{.16\textwidth}
			\resizebox{\textwidth}{!}{\input{./figure/email-Eu_p_n_b_38_10_5_25_Avg-Nbr-Degree-Weighted_2.pgf}}
			\vspace{-5mm}
			\caption{\footnotesize Eu ($|S|=2$)}
		\end{subfigure}
		\begin{subfigure}{.16\textwidth}
			\resizebox{\textwidth}{!}{\input{./figure/email-Eu_p_n_b_38_10_5_25_Avg-Nbr-Degree-Weighted_3.pgf}}
			\vspace{-5mm}
			\caption{\footnotesize Eu ($|S|=3$)}
		\end{subfigure}
		\begin{subfigure}{.16\textwidth}
			\resizebox{\textwidth}{!}{\input{./figure/email-Eu_p_n_b_38_10_5_25_Avg-Nbr-Degree-Weighted_4.pgf}}
			\vspace{-5mm}
			\caption{\footnotesize Eu ($|S|=4$)}
		\end{subfigure}
		\caption{\label{fig:node_fig}\textbf{Node Features vs. Node Persistence.} 
		The distribution of $w$ (i.e., weighted degree) and $\bar{w}$ (i.e., the average weighted degree of neighbors) of nodes with each level of $k$-node persistence in two datasets. \textcolor{blue}{Blue} lines indicate means and \textcolor{orange}{orange} lines indicate medians. See \cite{appendix} for results in the other datasets.}
	\end{figure*}

\subsection{Local Analysis (3): Node Features vs. Node Persistence}
\label{sec:obs:node}
	
	Below, we explore the relations between the structural features of each node\footnote{For each node $v$, let $t_v$ be the time when $v$ is involved in any HOI of size $k$ for the first time.
	Then, we measure the structural node features of $v$ using only the hyperedges appearing between time $t_v+1$ and $t_v+T_s$. We set $T_s$ to $5$ as in Section~\ref{sec:obs:defn}.} and its $k$-node persistence.
	The \textit{$k$-node persistence} of a node is the average persistence of the HOIs of size $k\in\{2,3,4\}$ that the node is involved in. 
	As the structural node features, we use those defined in Section~\ref{sec:obs:node_group}.
	
	
	\smallsection{Observations.}
	We report in Table~\ref{tab:feature_persistence_tab} the mutual information (MI) and Pearson correlation coefficients (CC) between each structural node feature and the $k$-node persistence.
	The MIs are at least $0.21$ for every node feature. 
	On average, the MI is largest for $r$ (i.e., PageRank), followed by $\bar{w}$ (i.e., the average weighted degree of neighbors), and then $\bar{d}$ (i.e., the average degree of neighbors).
	The correlation is strongest for $o$ (i.e., the number of occurrences) and $w$ (weighted node degree), which are positively correlated with the $k$-node persistence.
	Among the features, only $\bar{w}$, $\bar{d}$, and $l$ (i.e., the local clustering coefficient) are negatively correlated with the $k$-node persistence.
	The distributions of $w$ and $\bar{w}$ of nodes with each level of $k$-node persistence in two datasets are shown in Fig.~\ref{fig:node_fig}.
	
\vspace{0.5mm}
	\noindent\fbox{%
    \parbox{0.98\columnwidth}{
    \vspace{-3.5mm}
        \begin{obs}\label{obs:node_node}
            \textsc{(Node Features vs. Node Persistence)}
	        In real-world hypergraphs, the weighted degree and number of occurrences of each node are positively correlated with the persistence of HOIs that the node is involved in.
	    \end{obs}
	    \vspace{-3.5mm}
    }%
    }
    

    \section{Predictability and Predictors}
    \label{sec:predict}

In this section, we review our experiments for Q1-Q3: 
\begin{itemize}[leftmargin=*]
    \setlength\itemsep{0em}
    \item \textbf{Q1. Predictability}: How accurately can we predict the persistence of HOIs using the structural features?
    \item \textbf{Q2. Feature Importance}: Which structural features are important in predicting the persistence?
    \item \textbf{Q3. Effect of Observation Periods:} How does the period of observation for measuring the structural features affect the prediction accuracy? 
\end{itemize}

\subsection{Experimental Settings}
    
    Below, we describe the experimental settings. See Section~\ref{sec:prelim:data} for the datasets.

    \smallsection{Problem Formulation.}
    We consider Problems~\ref{problem:group} and \ref{problem:node} and formulate them as regression problems.
    As earlier, we assume $T_{s}=5$ and $T_{p}=10$, unless otherwise stated. Thus, both the persistence of HOIs and the $k$-node persistence of nodes lie between $0$ and $10$.
    
\vspace{1mm}
	\noindent\fbox{%
        \parbox{0.98\columnwidth}{%
        \vspace{-4mm}
        \begin{problem}[Persistence Prediction] \label{problem:group}
        \hfill
        \vspace{-3.5mm}
        \begin{itemize}[leftmargin=*]
            \item \textbf{Given:} 
             \begin{itemize}[leftmargin=*]
                \item \vspace{-3mm} a HOI $S$ appearing for the first time at time $t$,
                \item \vspace{-1.5mm} all hyperedges appearing between time $t$ and $t+T_s$,
             \end{itemize}
            \item \vspace{-2mm} \textbf{Predict:} the persistence of $S$ between time $t+T_s+1$ and $t+T_s+T_p$.
        \end{itemize}
        \end{problem}
        \vspace{-3.5mm}
        }
    }
    \vspace{1mm}
    
    \vspace{1mm}
	\noindent\fbox{%
        \parbox{0.99\columnwidth}{%
        \vspace{-4mm}
        \begin{problem}[$k$-Node Persistence Prediction] \label{problem:node}
        \hfill
        \vspace{-3.5mm}
        \begin{itemize}[leftmargin=*]
            \item \textbf{Given:} 
             \begin{itemize}[leftmargin=*]
                \item \vspace{-3mm} a node $v$ involved in a HOI of size $k$ for the first time at time $t$,
                \item \vspace{-1.5mm} all hyperedges appearing between time $t$ and $t+T_s$,
             \end{itemize}
            \item \vspace{-2mm} \textbf{Predict:} the $k$-node persistence of $v$.
        \end{itemize}
        \end{problem}
        \vspace{-3.5mm}
        }%
    }
    

    \smallsection{Prediction Methods.}
    We use the $8$ group features and the $8$ node features (see Section~\ref{sec:obs}) as the input features, and they are measured on the hypergraph consisting of the given hyperedges (see the problem definitions).
    We consider four regression models: \textit{multiple linear regression} (\textbf{LR}), \textit{random forest regression} (\textbf{RF}), \textit{linear support vector regression} (\textbf{SVR}), and \textit{multi-layer perceptron regressor} (\textbf{MLP}).
    RFs have $30$ decision trees with a maximum depth of $10$ and MLPs have one hidden layer with $(2\times(\text{the number of features used}) + 1)$ neurons and tanh as the activation function.
    We also consider the \textbf{mean} ($k$-node) persistence in the training set as baseline. 
    In each hypergraph, we use $2/3$ of the HOIs (and their persistence) and $4/5$ of the nodes (and their $k$-node persistence) for training; and we use the remaining ones for testing.
    See \cite{appendix} for the distribution of persistence in each hypergraph.
    
    \smallsection{Evaluation Methods.}
    We evaluate the predictive performance of the regression models using two metrics: \textit{coefficients of determination} ($\mathbf{R^2}$), which measures how well the predictions approximate the real data, and \textit{root mean squared error} (\textbf{RMSE}).
    A higher $R^2$ and lower RMSE indicate better performance.


\subsection{Q1. Predictability}
	\label{sec:exp:predictability}
    The predictive performance of the regression models is summarized in Table~\ref{tab:predictability_results}.
    For both tasks, \textbf{RF} was most accurate. On average, compared to baseline, \textbf{RF} reduces RMSE $\mathbf{46.5\%}$ and $\mathbf{27.8\%}$ for Problems~\ref{problem:group} and \ref{problem:node}, respectively.  
    The performance gap between \textbf{RF} and baseline grows as the size of HOIS (i.e., $|S|$) increases.
    For example, for Problem~\ref{problem:group}, the gap is $\mathbf{38.1\%}$, $\mathbf{50.0\%}$, and $\mathbf{64.8\%}$ when the sizes of HOIs are $2$, $3$ and $4$, respectively. 
    
	\noindent\fbox{%
        \parbox{0.98\columnwidth}{%
        \vspace{-3.5mm}
        \begin{obs}\label{obs:predictability} \textsc{(Predictability and the Size of HOIs)} In real-world hypergraphs,
        the structural features are useful for predicting the persistence of HOIs and the $k$-node persistence of nodes, especially when the size of the HOI is large.
        \end{obs}
        \vspace{-3.5mm}
        }%
    }
    
    \begin{table}[t]
        \vspace{-5mm}
        \small
    	\caption{\textbf{Predictability.} For each task, the first and second most accurate models are in \textbf{bold} and \underline{\smash{underlined}}, respectively. Using the structural features, all regression models achieve much better predictive performance than baseline (i.e., \textbf{mean}), especially when the size of the HOI (i.e., $|S|$) is large.
    	\textbf{RF} performs best.
    	}\label{tab:predictability_results}
    	
        	\centering
                \begin{tabular}{c|ccc|ccc|ccc|ccc}
                \toprule
                Target                                                       & \multicolumn{6}{c|}{Persistence of HOIs}                                                                        & \multicolumn{6}{c}{$k$-Node Persistence of Nodes}                                                                                    \\ \midrule
                Measure                                                & \multicolumn{3}{c|}{$R^2$*}                                      & \multicolumn{3}{c|}{RMSE** }                                     & \multicolumn{3}{c|}{$R^2$}                                                           & \multicolumn{3}{c}{RMSE}                                      \\ \midrule
                Size of HOIs \ & 2             & 3             & 4             & 2             & 3             & 4             & 2             & 3             & \multicolumn{1}{c|}{4}             & 2             & 3             & 4             \\ \midrule
                Mean                                                   & 0.00          & 0.00          & 0.00          & 1.29          & 0.73          & 0.60          & -0.01         & -0.01         & \multicolumn{1}{c|}{-0.03}         & 0.75          & 0.56          & 0.54          \\ \midrule
                SVR                                                    & 0.17          & 0.13          & 0.10          & 1.12          & 0.63          & 0.48          & 0.03          & 0.01          & \multicolumn{1}{c|}{0.00}          & 0.73          & 0.56          & 0.54          \\
                LR                                                     & 0.28          & 0.22          & 0.23          & 1.05          & 0.58          & 0.45          & {\ul 0.17}          & {\ul 0.15}          & \multicolumn{1}{c|}{{\ul 0.09}}    & {\ul 0.75}    & {\ul 0.71}    & {\ul 0.67}    \\
                MLP                                                    & {\ul 0.34}    & {\ul 0.31}    & {\ul 0.37}    & {\ul 0.95}    & {\ul 0.53}    & {\ul 0.42}    & 0.14    & 0.06    & \multicolumn{1}{c|}{0.02}          & 0.77          & 0.75          & 0.72          \\
                \textbf{RF}                                            & \textbf{0.61} & \textbf{0.62} & \textbf{0.68} & \textbf{0.83} & \textbf{0.38} & \textbf{0.24} & \textbf{0.61} & \textbf{0.66} & \multicolumn{1}{c|}{\textbf{0.71}} & \textbf{0.54} & \textbf{0.41} & \textbf{0.39} \\ \bottomrule
                \multicolumn{12}{l}{$*$The higher, the better. $**$The lower, the better.} \\
                \end{tabular}
	\end{table}
    
     \begin{table*}[t]
	        \vspace{-2mm}
            \centering
        	\caption{\label{tab:feature_rank}\textbf{Feature Importance.} We report the ranking of the importance of each feature. 
        	We use Gini importance based on \textbf{RF} for feature importance, and we average the rankings over all $13$ hypergraphs.
        	For each task, the first, second, and third most important features are in \textbf{bold}, \underline{\smash{underline}}, and \textit{italic}, respectively.}
            \setlength{\tabcolsep}{5pt}
            \scalebox{0.78}{
                \begin{tabular}{c|cccccccc|cccccccc|cccccccc}
                \toprule
                \multicolumn{1}{l|}{}                                  & \multicolumn{16}{c|}{Prediction of Persistence of HOIs}                                                                                                                                                                                                                                                                                                                                                      & \multicolumn{8}{c}{Prediction of $k$-Node Persistence of Nodes}                                                                                    \\ \midrule
\begin{tabular}[c]{@{}c@{}}Size\\ of HOIs\end{tabular} & \textbf{$\#$} & \textbf{$\frac{\#}{\cup}$} & \textbf{$\frac{\Sigma}{\Sigma\cup}$} & \textbf{$\cap$} & \textbf{$\frac{\#}{\cap}$} & \textbf{$\frac{\Sigma}{\cap}$} & \textbf{$\frac{\Sigma}{\#}$} & $H$          & \textbf{$d$} & \textbf{$w$} & \textbf{$c$} & \textbf{$r$} & \textbf{$\bar{d}$} & \textbf{$\bar{w}$} & \textbf{$l$} & \textbf{$o$} & \textbf{$d$} & \textbf{$w$} & \textbf{$c$} & \textbf{$r$} & \textbf{$\bar{d}$} & \textbf{$\bar{w}$} & \textbf{$l$} & \textbf{$o$} \\ \midrule
2                                                      & \textbf{2.8}  & 10.7                       & 8.6                                  & 13.1            & 13.3                       & 9.0                            & 9.2                          & 8.7          & 9.9          & 8.6          & 8.8          & 5.9          & \textit{4.9}       & {\ul 4.3}          & 6.4          & 11.9         & 6.7          & 4.3          & 7.2          & {\ul 3.2}    & 3.4                & \textbf{2.9}       & 5.3          & {\ul 3.2}    \\
3                                                      & \textit{5.4}  & 9.2                        & 9.2                                  & 11.8            & 11.2                       & 9.6                            & 9.8                          & 7.9          & 11.2         & 9.1          & 8.4          & 5.7          & {\ul 5.1}          & \textbf{4.3}       & 6.4          & 12.0         & 6.6          & 4.1          & 7.3          & \textbf{2.7} & \textit{3.5}       & \textbf{2.7}       & 5.0          & 4.3          \\
4                                                      & \textbf{5.3}  & 9.3                        & 9.9                                  & 10.3            & 10.6                       & 8.3                            & 8.7                          & \textit{7.0} & 9.5          & 7.3          & 9.2          & 7.7          & 7.7                & {\ul 6.3}          & 8.0          & 11.0         & 6.1          & 4.0          & 6.6          & \textbf{2.6} & \textit{3.5}       & {\ul 3.1}          & 5.3          & 4.9          \\ \midrule
Avg                                                    & \textbf{4.5}  & 9.7                        & 9.2                                  & 11.7            & 11.7                       & 9.0                            & 9.2                          & 7.9          & 10.2         & 8.3          & 8.8          & 6.4          & \textit{5.9}       & {\ul 5.0}          & 6.9          & 11.6         & 6.4          & 4.1          & 7.0          & \textbf{2.8} & \textit{3.5}       & {\ul 2.9}          & 5.2          & 4.1          \\ \bottomrule
                \end{tabular}
            }
    	\end{table*}
	

\subsection{Q2. Feature Importance}
    \label{sec:exp:feature}
    In this subsection, we analyze the usefulness of the structural features in predicting the persistence of HOIs and the $k$-node persistence of nodes.
    Recall that we examine the relevance between ($k$-node) persistence and each individual feature independently in Section~\ref{sec:obs}.
    Here, we take the potential correlation between the features, which is previously ignored, into consideration. 
    As inferred from the fact that \textbf{LR} (i.e,. multiple linear regression) is significantly and consistently outperformed by \textbf{RF}, the \textbf{LR} suffers from multicollinearity and/or the linearity assumption, which \textbf{LR} is based on, is clearly invalid for Problems~\ref{problem:group} and \ref{problem:node}.
    Thus, while \textbf{LR} is a standard way of measuring the significance of multiple features at the same time, here, we employ feature analysis methods based on \textbf{RF}, which is consistently most accurate for both problems. The results of \textbf{LR} can be found in \cite{appendix}.  
    
    
    \begin{figure*}[t]
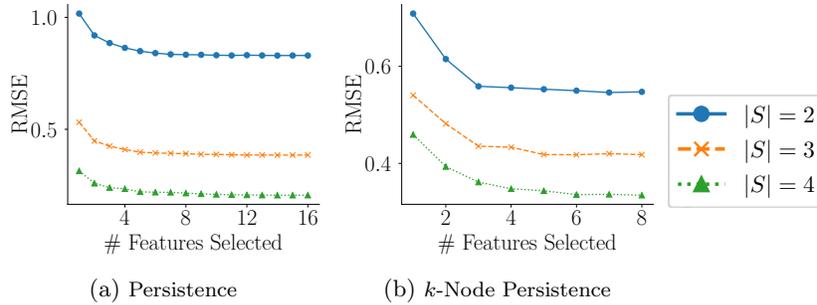

        \vspace{-6mm}
        \centering
		\captionsetup[subfigure]{justification=centering}
			\begin{subfigure}{.25\textwidth}
				\resizebox{\textwidth}{!}{\input{./figure/t_b_s_10_5_25.pgf}}
				\caption{\footnotesize Persistence} \label{fig:feature_selection_fig_1}
			\end{subfigure}
			\begin{subfigure}{.25\textwidth}
				\resizebox{\textwidth}{!}{\input{./figure/t_b_s_10_5_25_2.pgf}}
				\caption{\footnotesize $k$-Node Persistence} \label{fig:feature_selection_fig2_1}
			\end{subfigure}
			\begin{subfigure}{.12\textwidth}
		        \centering
				\resizebox{\textwidth}{!}{\input{./figure/t_b_s_10_5_25_legend.pgf}}
		    \end{subfigure}
			\caption{\label{fig:feature_selection_fig}\textbf{Feature Selection.} We show how the RMSE of \textbf{RF} changes as we remove the input features one by one based on their importance.}  
	\end{figure*}
    
	\smallsection{Which Features are Useful?}
	We first describe the \textit{Gini importance}, which we use to measure the importance of each structural feature.
	A trained random forest regressor (\textbf{RF}) consists of multiple decision trees, each of which has multiple internal nodes.
	In each internal node, a feature is used to divide data into separate sets, and the importance of the feature can be measured by the reduction of the variance of ($k$-node) persistence due to the feature.
	We measure how much each feature reduces the variance in each tree, and we use the average over all trees as the final importance of each feature.	
		
	We measure the importance of each structural feature as described above in each dataset, and based on the importance, compute the rankings of the features.
	In Table~\ref{tab:feature_rank}, we average the rankings of the features over all $13$ datasets, instead of averaging the importance itself, whose scale varies in different datasets.
	
	\vspace{0.5mm}
	\noindent\fbox{%
        \parbox{0.98\columnwidth}{%
        \vspace{-3.5mm}
        \begin{obs}\label{obs:importance:group}\textsc{(Strong Predictors for Problem~\ref{problem:group})} In predicting the persistence of each HOI $S$ in
        real-world hypergraphs, the number of hyperedges containing $S$ (i.e., \#), and the average (weighted) degree of the neighbors of each node in $S$ (i.e., $\bar{w}$ and $\bar{d}$) are most useful among the $16$ features.
        \end{obs}
        \vspace{-3.5mm}
        }%
    }
    \vspace{1mm}
	    
	\vspace{1mm}
	\noindent\fbox{%
        \parbox{0.98\columnwidth}{%
        \vspace{-3.5mm}
        \begin{obs}\label{obs:importance:node}\textsc{(Strong Predictors for Problem~\ref{problem:node})} In predicting the $k$-node persistence of each node in
    real-world hypergraphs, its PageRank (i.e., $r$) and the average (weighted) degree of its neighbors (i.e., $\bar{w}$ and $\bar{d}$) are most useful among the $8$ features.
        \end{obs}
        \vspace{-3.5mm}
        }%
    }
    

    	\smallsection{How Many Features are Needed?}
		Based on the feature importance measured as described above (i.e., Gini importance), we optimize the number of input features in the regression models for Problems~\ref{problem:group} and \ref{problem:node}.
		Specifically, we start from all the features and repeat removing one feature with the lowest feature importance.
		Whenever a feature is removed, we measure the RMSE of \textbf{RF}, which consistently performs best (see Table~\ref{tab:predictability_results}).
	    We report in Fig.~\ref{fig:feature_selection_fig} the average RMSE over all $13$ hypergraphs. 
	    We observe clear diminishing returns. That is, the reduction of RMSE due to an additional feature decreases as we have more.
	    For both problems, using about a half of the considered input features (i.e., $8$ features for Problem~\ref{problem:group} and $4$ features for Problem~\ref{problem:node}) yields sufficiently small RMSE, and the amount of improvement from additional features is negligible.
	    
\vspace{0.5mm}
	\noindent\fbox{%
        \parbox{0.98\columnwidth}{%
        \vspace{-3.5mm}
        \begin{obs}[Proper Number of Features] \label{obs:feature_num} 
        In predicting the persistence of HOIs and the $k$-node persistence of nodes in real-world hypergraphs, using about a half of the considered structural features based on their importance yields similar 
        predictive performance, compared to using all the features.
        \end{obs}
        \vspace{-3.5mm}
        }%
    }   
	    

\subsection{Effect of Observation Periods}
	
    We investigate how the length of the observation period (i.e. $T_s$) during which we measure structural features affects the predictability of the persistence of HOIs and the $k$-node persistence of nodes.
    For both Problems~\ref{problem:group} and \ref{problem:node}, we measure the RMSE of \textbf{RF}, which consistently performs best (see Table~\ref{tab:predictability_results}), and its improvement in percentage over baseline (i.e, \textbf{Mean}), as we increase $T_s$ from $1$ to $5$.
    Note that, in all settings, we measure ($k$-node) persistence over $T_{p}=10$ time units right after the observation period.
    We also measure the effect of lengthening the observation period so that it additionally includes the period before each considered HOI appears for the first time.
    This change is equivalent to using additionally all hyperedges appearing before the first appearance of each considered HOI when measuring structural features.
    
    As seen in Table~\ref{tab:performance_analysis}, when predicting the persistence of HOIs, the predictive performance gets better, and the improvement over baseline grows, as the observation period increases.
    However, there is no such tendency when predicting the $k$-node persistence of nodes.
    
\vspace{0.5mm}
    	\noindent\fbox{%
        \parbox{0.98\columnwidth}{%
        \vspace{-3.5mm}
        \begin{obs}\label{obs:period}
        \textsc{(Effect of Observation Periods)}
        Observing HOIs in real-world hypergraphs for longer periods of time enables us to better predict their persistence in the future.
       \end{obs}
       \vspace{-3.5mm}
        }%
    }   
    
    \begin{table}[t]
        \vspace{-2mm}
        \small
	    \centering
        \caption{\label{tab:performance_analysis}\textbf{Effects of Observation Periods.} We report the RMSE of \textbf{RF} and its improvement in percentage over the baseline method (i.e., \textbf{Mean}) under different settings.
        The column `Past' indicates whether the observation period additionally includes the period before each considered HOI appears for the first time (`O') or not (`X').
        For each task, the best and second best results are in \textbf{bold} and \underline{\smash{underlined}}, respectively.
        }
            \begin{tabular}{c|c|ccc|ccc|ccc|ccc}
            \toprule
            \multicolumn{2}{c|}{Target} & \multicolumn{6}{c|}{Persistence of HOIs}                                                                            & \multicolumn{6}{c}{$k$-Node Persistence of Nodes}                                                                  \\ \midrule
            \multicolumn{2}{c|}{Measure}  & \multicolumn{3}{c|}{RMSE* of RF}                             & \multicolumn{3}{c|}{Improvement  (in \%)}                     & \multicolumn{3}{c|}{RMSE  of RF}                             & \multicolumn{3}{c}{Improvement (in \%)}                   \\ \midrule
            $T_s$                & Past                  & 2**             & 3             & 4             & 2              & 3              & 4              & 2             & 3             & 4             & 2              & 3              & 4               \\ \midrule
            1                    & X                     & 0.96          & 0.48          & \multicolumn{1}{c|}{0.32}          & 31.6          & 42.3          & 50.7          & 0.62          & 0.46          & 0.43          & 18.5          & 25.5          & 31.8          \\
            3                    & X                     & 0.88          & 0.42          & \multicolumn{1}{c|}{0.28}          & 34.1          & 45.4          & 55.0          & 0.55          & \textbf{0.41} & \textbf{0.38} & 24.8          & \textbf{29.1} & \textbf{34.4} \\
            5                    & X                     & {\ul 0.83}    & {\ul 0.38}    & \multicolumn{1}{c|}{{\ul 0.24}}    & {\ul 36.0}    & {\ul 47.7}    & {\ul 59.4}    & {\ul 0.54}    & 0.41          & 0.39          & {\ul 27.4}    & 26.4          & 27.5          \\ \midrule
            1                    & O                     & 0.95          & 0.47          & \multicolumn{1}{c|}{0.30}          & 32.5          & 42.5          & 53.6          & 0.62          & 0.46          & 0.45          & 19.2          & 24.9          & 28.7          \\
            3                    & O                     & 0.87          & 0.41          & \multicolumn{1}{c|}{0.27}          & 35.0          & 46.4          & 57.0          & 0.55          & 0.42          & {\ul 0.38}    & 25.0          & {\ul 28.2}    & {\ul 34.2}    \\
            5                    & O                     & \textbf{0.81} & \textbf{0.38} & \multicolumn{1}{c|}{\textbf{0.23}} & \textbf{37.2} & \textbf{48.6} & \textbf{60.8} & \textbf{0.54} & {\ul 0.41}    & 0.39          & \textbf{28.2} & 26.5          & 27.9          \\ \bottomrule
            \multicolumn{12}{l}{$*$The lower, the better. $**$The size of HOIs (i.e., $|S|$).} \\
            \end{tabular}
	\end{table}

    \section{Conclusions}
    \label{sec:summary}
    In this study, we examine the persistence of high-order interactions (HOIs) at hypergraph-, group-, and node-levels in $13$ real-world hypergraphs using $16$ structural features.
Our main findings are summarized as follows: 

\begin{itemize}[leftmargin=*]
    \setlength\itemsep{0em}
    \item 
    The persistence of HOIs follows power-law distributions whose exponents drop as the sizes of HOIs grow.
    \item 
    Some structural features of HOIs (e.g., entropy in the sizes of hyperedges including them) are closely related to their persistence.
    \item 
    The structural features are informative, leading to accuracy gains in forecasting the persistence of HOIs.
    \item The accuracy gains get larger as (a) the size of HOIs grows and (b) we observe HOIs for a longer period.
    \item The strongest predictors of the persistence of a HOI are (a) the number of hyperedges containing the HOI and (b) the average (weighted) degree of the neighbors of each node in the HOI.
\end{itemize}
\textbf{Reproducibility:} The source code and datasets used in the paper are available at \cite{appendix}.

\smallsection{Acknowledgements} This work was supported by National Research Foundation of Korea (NRF) grant funded by the Korea government (MSIT) (No. NRF-2020R1C1C1008296),
and Institute of Information \& Communications Technology Planning \& Evaluation (IITP) grant funded by the Korea government (MSIT) (No. 2019-0-00075, Artificial Intelligence Graduate School Program (KAIST)).

    \vspace{-2mm}
    \bibliographystyle{plain}
    \bibliography{bibliography}

\end{document}